\documentclass[twocolumn,showpacs,aps]{revtex4}

\usepackage{graphicx}

\usepackage{times}

\def\hb{\hbar}

\def\O{\Omega}
\def\OO{\Omega/\omega_\perp}
\def\o{\omega}
\def\bO{{\bf \Omega}}
\def\tO{{{\tilde \Omega}}}
\def\tbO{{{\tilde {\bf \Omega}}}}
\def\ep{\varepsilon}

\def\O{\Omega}

\def\a0{\alpha_0}
\def\a{\alpha}
\def\b{\beta}

\def\rtf{\rho_{\hbox{\tiny TF}}}

\def\r0{\rho_{0}}

\def\be{\begin{equation}}
\def\ee{\end{equation}}
\def\beq{\begin{equation}}
\def\eeq{\end{equation}}
\def\ds{\displaystyle}

\def\cd{{\cal D}}

\begin{document}

\title{Giant vortices in combined harmonic and quartic traps}

\author{Amandine Aftalion}
\email{aftalion@ann.jussieu.fr} \affiliation{Laboratoire
Jacques-Louis Lions,  Universit{\'e} Paris 6, 175 rue du Chevaleret,
75013 Paris, France.}
\author{Ionut Danaila}
\email{danaila@ann.jussieu.fr} \affiliation{Laboratoire
Jacques-Louis Lions,  Universit{\'e} Paris 6, 175 rue du Chevaleret,
75013 Paris, France.}
\date{\today}


\begin{abstract}
We consider a rotating Bose-Einstein condensate confined in 
combined harmonic and quartic traps, following recent experiments
[V. Bretin, S. Stock, Y. Seurin and J. Dalibard,
cond-mat/0307464].
We investigate numerically
the behavior of the wave function which solves the
three-dimensional Gross Pitaevskii equation. When the harmonic
part of the potential is dominant, as the angular velocities $\O$ increases, the vortex lattice evolves into a
giant vortex. We also investigate a case not covered by the experiments or the previous numerical works: for
 strong quartic potentials, the giant vortex is obtained
for lower $\O$, before the lattice is formed. We analyze in detail
the three dimensional structure of vortices.
 \end{abstract}

\maketitle

\section{Introduction}

The existence and formation of quantized vortices have recently
been widely studied  in Bose Einstein condensates \cite{mal,
MCWD2,MCWD,RK,AK,RBD}. One type of experiments consists in
rotating the magnetic trap confining the atoms. For a harmonic
trapping potential $(1/2)m\o_\perp^2r^2$, and a rotating frequency
$\O$ close to $0.7 \o_\perp$, vortices start to appear and arrange
themselves into a lattice \cite{MCBD}.  As $\O$ is
 increased, the number of vortices increases as well.
 In the case of a harmonic
trap, the confinement and the centrifugal force prevent the
condensate from rotating at a frequency $\O$ beyond $\o_\perp$.
 The regime of fast rotation  is especially
interesting since it provides a setting for a large number of
vortices and eventually giant vortices \cite{Fi,E}.

Theoretical and numerical studies have considered stiffer
 potentials than the harmonic one, behaving like $r^n$ or $r^2+r^4$
  \cite{F,L,K,Kav}. This type of trapping,
which eliminates the singular behavior at $\O=\o_\perp$, has
recently been achieved experimentally  by superimposing  a blue
detuned laser beam to the magnetic trap holding the atoms
\cite{BD}. The resulting potential is \beq\label{trap}
V_{trap}(r,z)=V_{h}(r,z)+U(r),
\eeq with \beq\label{pot-exp} V_{h}={1\over 2}m
  \omega^2_{\perp} r^2+{1\over 2}m
  \omega^2_{z} z^2,
   \hbox{ and } U(r)=U_0 \exp\bigl ( -{{2r^2}\over{w^2}}\bigr ). \eeq
 For $r/w$ sufficiently small,   the
resulting potential can be approximated by :
\beq\label{pot-approx} V_{trap}\simeq \left[{1\over 2}m
  \omega^2_{\perp} -{{2U_0}\over w^2}\right] r^2 +{{2U_0}\over
  w^4}r^4 + {1\over 2}m
  \omega^2_{z} z^2.
  \eeq

The purpose of this paper is to find the stable states (vortex
lattice, vortex array with hole and giant vortices) of the
condensate with this type of trapping potential and to analyze
their  three-dimensional structure. We consider a case similar to
the experiments and previous theoretical settings, where
 the amplitude $U_0$ of the superimposed laser is small, so that the coefficient of the $r^2$ term is positive.
  But we are especially interested in the case where the laser beam has sufficiently large amplitude so that
\beq\label{pot-ineq} {1\over 2}m
  \omega^2_{\perp} < {{2U_0}\over w^2}.\eeq
This changes the sign of the harmonic part of the potential
(\ref{pot-approx}). The point  is that, this case of a {\em
quartic minus harmonic} potential allows to observe giant vortices
at lower angular velocities than previously and the structure of vortices is different.


\section{Numerical approach}

We consider a pure BEC of $N$ atoms confined in a
 trapping potential $V_{trap}$, rotating along the $z$ axis at angular
 velocity $\O$. The equilibrium of the system corresponds to local
 minima of the Gross-Pitaevskii energy in the rotating frame

\begin{eqnarray}
\label{BE} \ds {\cal E}(\phi) &=& \ds \int_{\cal D} {\hb^2
\over{2m}} |\nabla \phi|^2+{{\hb \bO}}\cdot (i\phi, \nabla
\phi\times {\bf x}) \nonumber\\  & &\ds + V_{trap}\, |\phi|^2
+{{N\over 2}
  g_{3D}} |\phi|^4
\end{eqnarray}
where $g_{3D}=4\pi\hb ^2a/m$ and the wave function $\phi$ is
normalized to unity $\int_{\cal D} |\phi|^2 =1$.

For numerical purposes, it is  convenient to rescale the variables
as follows: ${\bf r}={\bf x}/R$, $ u({\bf r})=R^{3/2} \phi({\bf
x})$, where $R=d/\sqrt{\ep}$ and
\begin{equation}
d=\left(\frac{\hb}{m\omega_\perp}\right)^{1/2}, \,\,
\ep=\left({{d}\over {8\pi Na}}\right)^{2/5}, \,\, \tO
={\O}/{(\ep\omega_\perp)}.
\end{equation}
In this scaling, the trapping potential (\ref{pot-approx}) becomes
 \beq\label{pot-num}
V=(1-\a)r^2+{1\over 4} k r^4 + \b^2 z^2,\eeq where
\beq\label{params} \a=\frac{4U_0}{m \o_\perp^2 w^2}, \quad k= 4\a
\left(\frac{R}{w}\right)^2, \quad \beta=\frac{\o_z}{\o_\perp}.\eeq
Note that we take $\omega_\perp$ (which is the frequency of the
original harmonic potential $V_h$), and not
$\omega_\perp\sqrt{|1-\a|}$, as a scaling frequency for $\O$.
 For numerical applications, we choose
 $\ep=0.02$, $\beta=\o_z/\o_\perp=1/7$, $k/\a=0.25$, which fit the experimental values of Ref. \cite{BD}.
In \cite{BD}, $\a=0.25$, but we will take bigger values since our aim
is to understand the influence of $\a$ when it gets bigger than 1.

Then, we use the dimensionless energy introduced in \cite{AD} \beq
\label{be3d} {E}(u) = H(u) - \tO L_z(u),\eeq where $H$ is the hamiltonian
 \beq\label{H}
H(u)=\int{1\over 2} |\nabla u|^2 +{1\over 2\ep^2}V |u|^2+{1\over
4\ep^2}|u|^4\;,\eeq and $L_z$ the angular momentum
axis \beq\label{Lz} L_z(u)=i \int \bar{u} \left( y\frac{\partial
u}{\partial x} - x \frac{\partial u}{\partial y}\right). \eeq

Using a hybrid Runge-Kutta-Crank-Nicolson  scheme described in
Ref. \cite{AD}, we compute critical points of ${E}(u)$ by solving
the norm-preserving imaginary time propagation of the
corresponding equation: \beq\label{eqstrong} \frac{\partial
u}{\partial t}-{1\over 2}\nabla^2 u + i(\tbO\times {\bf r}).\nabla
u=-{1\over {2\ep^2}}u(V+|u|^2)+\mu_\ep u, \eeq  where $\mu_\ep$ is the Lagrange multiplier for the
constraint $\int_{\cal D} |u|^2=1$ and with $u=0$ on
$\partial \cd$ and. Here, $\cd$ is a rectangular
domain containing  the condensate.  A typical simulation uses a domain $(x,y,z)\in [-2,2]\times[-2,2]\times[-2.8,
2.8]$ with a refined grid of $200\times 200 \times 140$ nodes,
which  is sufficient to achieve grid-independence for all
considered numerical experiments.

\begin{figure}[!h]
\centerline{\includegraphics[width=0.7\columnwidth]{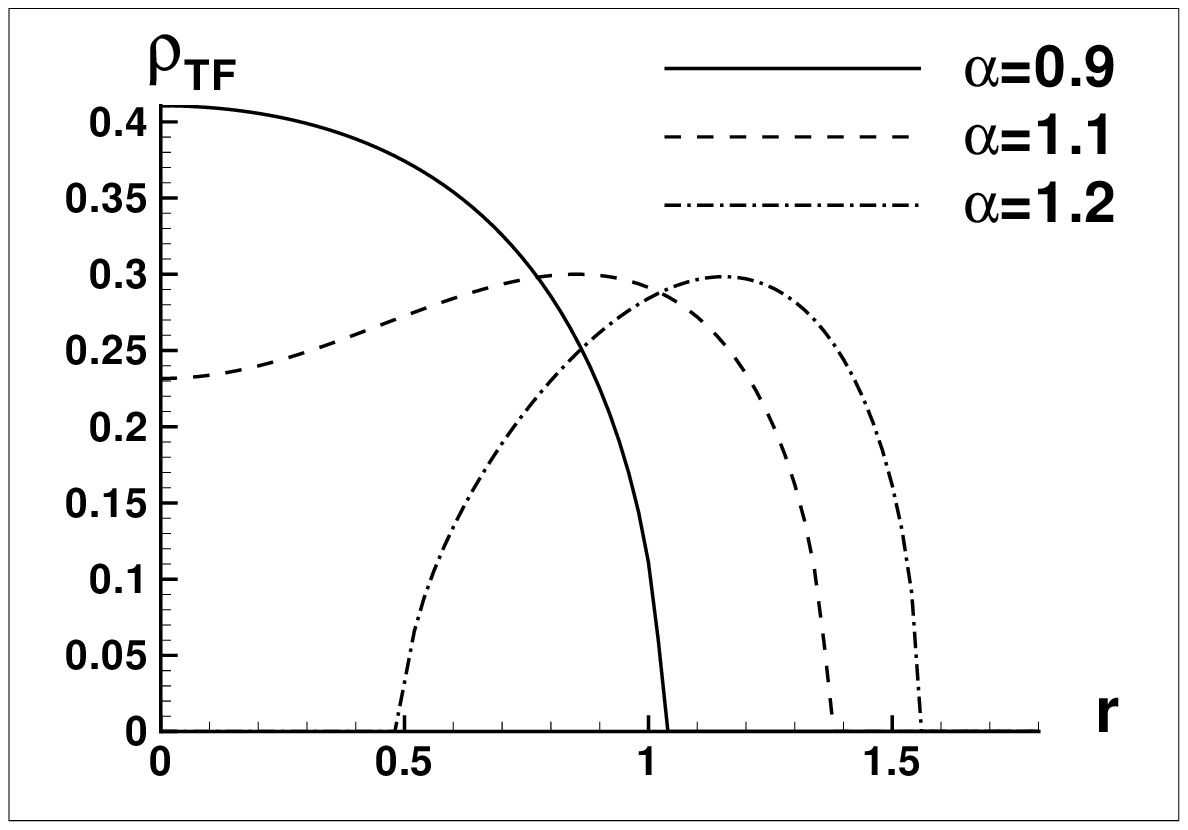}}
 \caption{Thomas-Fermi limit $\rtf$
 for different values of $\alpha$.} \label{fig-TFprof}
\end{figure}

\begin{figure}[!h]
\centerline{\includegraphics[width=0.90\columnwidth]{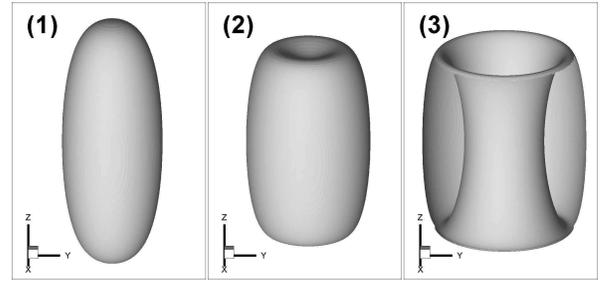}}
 \caption{Different shapes of the condensate at $\O=0$: isosurfaces of lowest density in the condensate for
 $\a=0.9$ (picture 1), $1.1$ (picture 2), $1.2$ (picture
 3).}
 \label{fig-cinit}
\end{figure}

We first compute the steady state corresponding to a nonrotating
($\O=0$) condensate, using as initial condition $u=\sqrt{\rtf}$,
 the Thomas-Fermi profile \beq\label{rtf} \rtf({\bf
r})=\r0 +(\a-1)r^2-{1\over 4} k r^4-\b^2z^2.\eeq Depending on the
choice of  $\alpha$, the Thomas-Fermi density profile can display
three different shapes, as shown in figure \ref{fig-TFprof}.
The corresponding steady solutions obtained for $\O=0$ (which will
be used as initial conditions for the subsequent runs with $\O>0$)
are displayed in figure \ref{fig-cinit}.
We can distinguish three cases:
\begin{itemize}
\item $\a<1$ (weak quartic case) is the case closest to the
experiments and is strongly influenced by the harmonic part. For
$\O=0$, a classical prolate condensate is obtained. As $\O$
increases, the effective trapping potential $ V^{eff}({\bf
r})=V({\bf r})-\ep^2\O^2r^2$ starts to have a mexican hat
structure. A vortex lattice appears for intermediate values of $\O$
and turns into a lattice with a hole for large $\O$. 
 \item $\a\gtrsim 1$ (intermediate quartic
case): the density profile has a depletion close to the center at
$\O=0$ but no hole.  The criterion for this case is
\beq\label{xi}\xi={{\sqrt \beta k^{5/4}}\over
{\pi(\a-1)^2}}>1.\eeq The density profile starts to have a hole
for intermediate values of $\O$.
 \item $\a> 1$  and $\xi<1$ (strong quartic case): the density profile has a hole for all
$\O$.
\end{itemize}

\section{Description of the results}

Depending on the values of $\a$ and $\O$, we observe different types of
configurations: vortex free configurations where the amplitude of
the wave function takes into account the  shape of the effective
trapping potential, vortex lattices,
 vortex arrays with hole and giant vortices.

\subsection{Intermediate quartic case ($\a=1.1$)}
\begin{figure}[!h]
\centerline{\includegraphics[width=0.90\columnwidth]{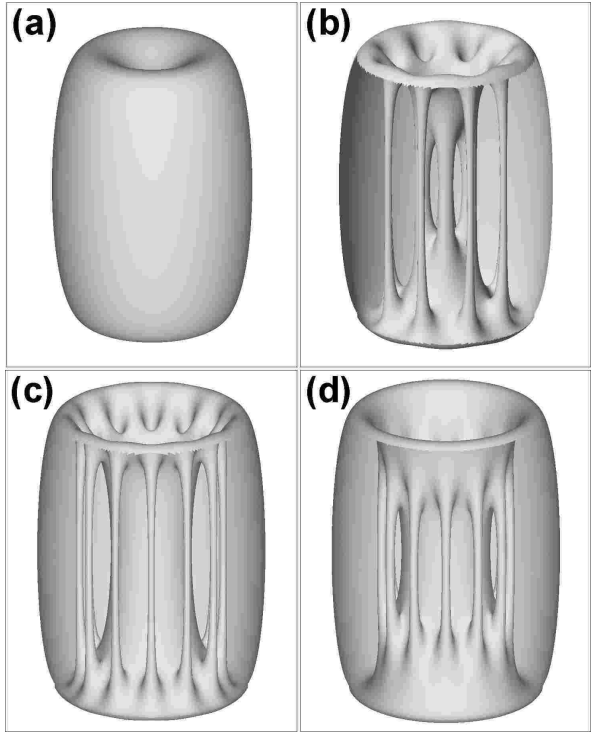}}
 \caption{ ($\a=1.1$) Side  view of the condensate  for
  ${\Omega/\o_\perp}=0.12$ (a), $0.2$ (b), $0.28$ (c),
   $0.32$ (d). Isosurface of lowest density.
} \label{fig-a11-side}
\end{figure}

\begin{figure}[!h]
\centerline{\includegraphics[width=0.90\columnwidth]{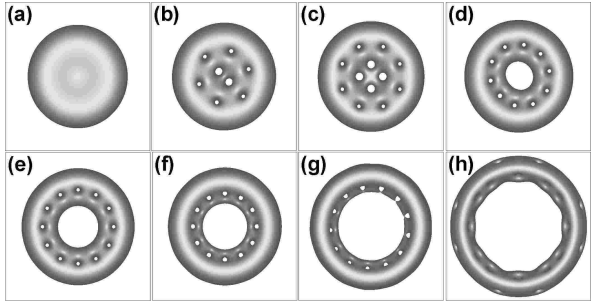}}
 \caption{ ($\a=1.1$) Top  view of the condensate  for
  ${\Omega/\o_\perp}=0.12$ (a), $0.16$ (b), $0.2$ (c), $0.24$ (d), $0.28$ (e),
   $0.32$ (f), $0.4$ (g) and $0.48$ (h). Isosurface of lowest density.
} \label{fig-a11-top}
\end{figure}

\begin{figure}[!h]
\centerline{\includegraphics[width=0.9\columnwidth]{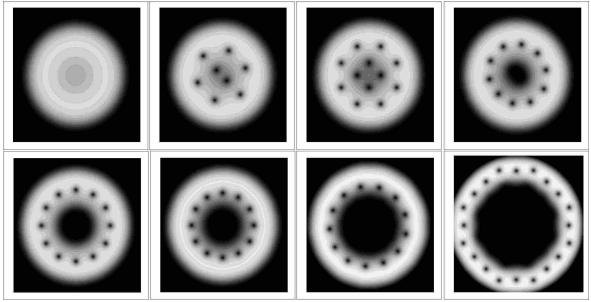}}
 \caption{($\alpha=1.1$)  Density contours in the plane $z=0$ for
  the same states as in figure \ref{fig-a11-top}: ${\Omega/\o_\perp}=0.12$, $0.16$, $0.2$, $0.24$, $0.28$,
   $0.32$, $0.4$ and $0.48$.} \label{fig-a11-2d}
\end{figure}

The potential $V$ has a Mexican hat structure. The isosurface of lowest density of the
solution is plotted in figure \ref{fig-a11-side}, the top view
in figure \ref{fig-a11-top} and in the middle plane $z=0$ in figure \ref{fig-a11-2d}. For $\O$ small, the density has a depletion
close to the center of the condensate but no hole and no vortices.
For $\O$ larger ($\OO\geq 0.16$), vortices are nucleated.

\begin{figure}[!h]
\centerline{\includegraphics[width=0.5\columnwidth]{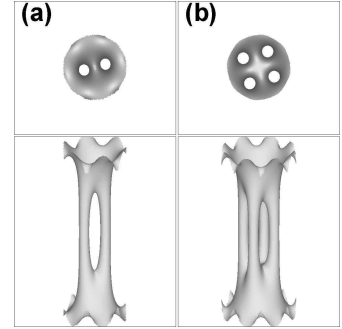}}
 \caption{ ($\a=1.1$) Vortex reconnection details for
  ${\Omega/\o_\perp}=0.16$ (a), $0.2$ (b).
} \label{fig3}
\end{figure}

For $0.16\leq \OO <0.24$,  the density of the solution is  zero
close to the top and bottom of the condensate, but not  at the
center,
 which gives rise to a special
structure of vortices: the vortices arrange themselves along two
concentric circles. The inner circle is made up of vortices which
are isolated in the center of the condensate but  reconnect
towards the top of the condensate (see the details in figure
\ref{fig3}). The outer circle is made up of almost straight $U$
vortices
 that reconnect to the inner circle
 close to the top and bottom of the condensate. As $\O$ increases, the number of
vortices on each circle increases. In figure \ref{fig-a11-top}(b),
the inner vortices seem to be bigger, but this is just an effect
due to the projection and the bending: the view at $z=0$ (figure
\ref{fig-a11-2d}) allows to check that all vortices have the same
size.

For $ \OO \geq 0.24$, the density profile of the solution is zero
in the center of the condensate, hence this creates a giant
vortex: the straight vortices that were close to the center on the
inner circle have merged into a giant vortex. There are also
isolated vortices regularly scattered on a circle around the giant
vortex. As $\O $ increases, the number of vortices inside and
outside the giant vortex increases and the length of the isolated
vortices decreases as can be seen in figure \ref{fig-a11-side}.

Note that the isolated vortices are
$U$ vortices that reconnect to the giant vortex at the center, not
to the boundary of the condensate, as in the case of the harmonic
trapping \cite{AD}, that is their bending is concave not convex.

\begin{figure}[!h]
\centerline{\includegraphics[width=0.60\columnwidth]{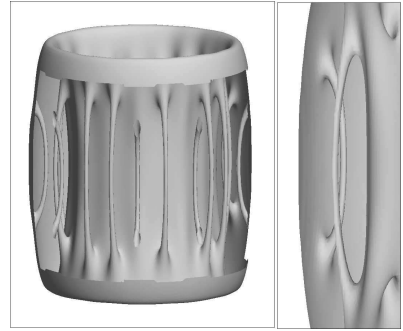}}
 \caption{ ($\a=1.1$) Vortex details for
  ${\Omega/\o_\perp}=0.48.$
} \label{fig4}
\end{figure}

For $\OO=0.48$, (see figure \ref{fig4}) the number of vortices has increased and there are
2 outer circles of vortices around the giant vortex: one circle of
$U$ vortices that reconnect to the giant vortex and one circle of
vortices that reconnect to the outer boundary of the condensate.
Both have different concavity in their bending as illustrated in
figure \ref{fig4}.

\subsection{Strong quartic potential case ($\a=1.2$)}

\begin{figure}[!h]
\centerline{\includegraphics[width=0.90\columnwidth]{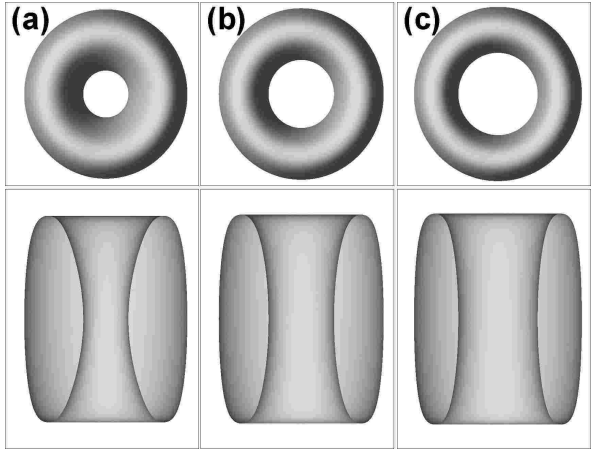}}
 \caption{ ($\a=1.2$) Top and side view of the condensate  for
  ${\Omega/\o_\perp}=0$ (a), $0.12$ (b) and $0.2$ (c).
} \label{fig5}
\end{figure}

The effective potential has a Mexican hat structure for all $\O$
and the density profile of the solution  always has a hole in the
center as illustrated in figure \ref{fig5}.

\begin{figure}[!h]
\centerline{\includegraphics[width=0.70\columnwidth]{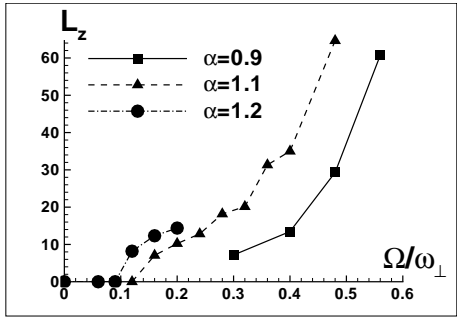}}
 \caption{Angular momentum $L_z$ (in units of $\hbar$) for all  studied configurations.
} \label{fig-lz}
\end{figure}

 For small $\O$, there are no vortices, that is $L_z=0$; it is only
the modulus of the solution that goes to zero. For larger $\O$
($\O/\o_\perp\ge 0.12$), the hole contains a giant vortex and
$L_z$ increases with $\O$ (see figure \ref{fig-lz}). We have not
found any isolated vortex around the giant vortex: all vortices
are included in the central giant vortex because of the strong
potential.
\begin{figure}[!h]
\centerline{\includegraphics[width=0.90\columnwidth]{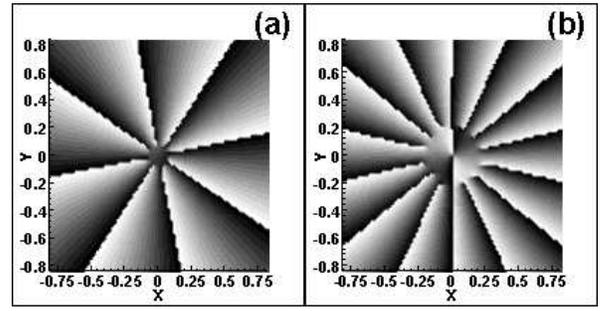}}
 \caption{($\a=1.2$) Phase distribution in a central $z=0$ cut plane.
  ${\Omega/\o_\perp}= 0.12$ (a) and $0.2$ (b).
} \label{fig-a12-phase}
\end{figure}
The giant vortex phase profiles (figure \ref{fig-a12-phase}) show
that the phase singularities do not completely overlap in the
center of the vortex. This feature has already been observed in
two-dimensional numerical simulation of a fast rotating condensate
by Kasamatsu {\em et al} \cite{K}. They described the giant vortex
as the hole containing single quantized vortices with such low
density that they are discernible only by the phase defects.


\subsection{Weak quartic case ($\a=0.9$)}
\begin{figure}[!h]
\centerline{\includegraphics[width=0.9\columnwidth]{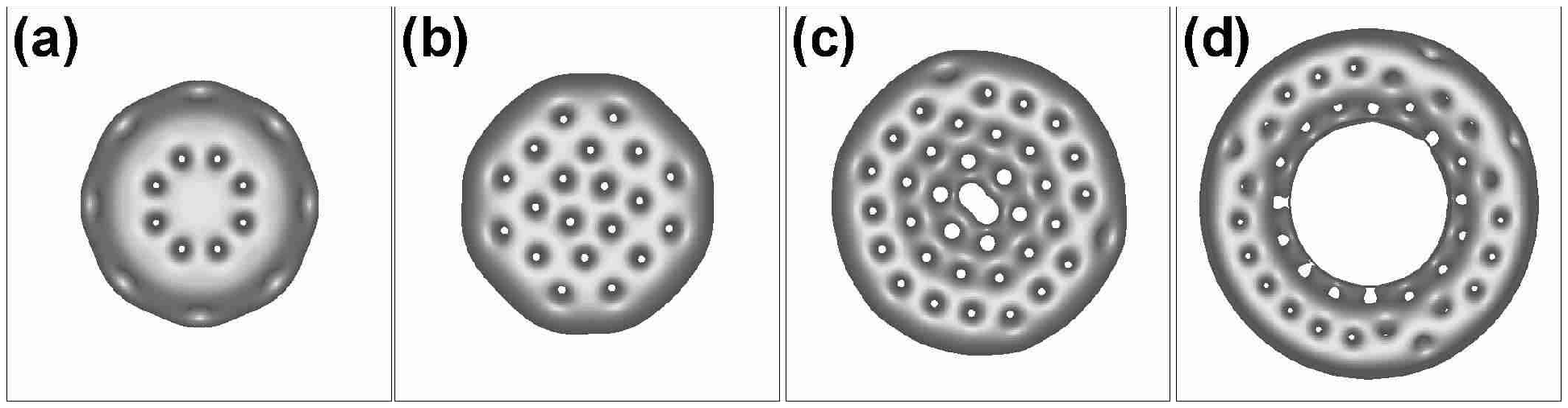}}
\centerline{\includegraphics[width=0.9\columnwidth]{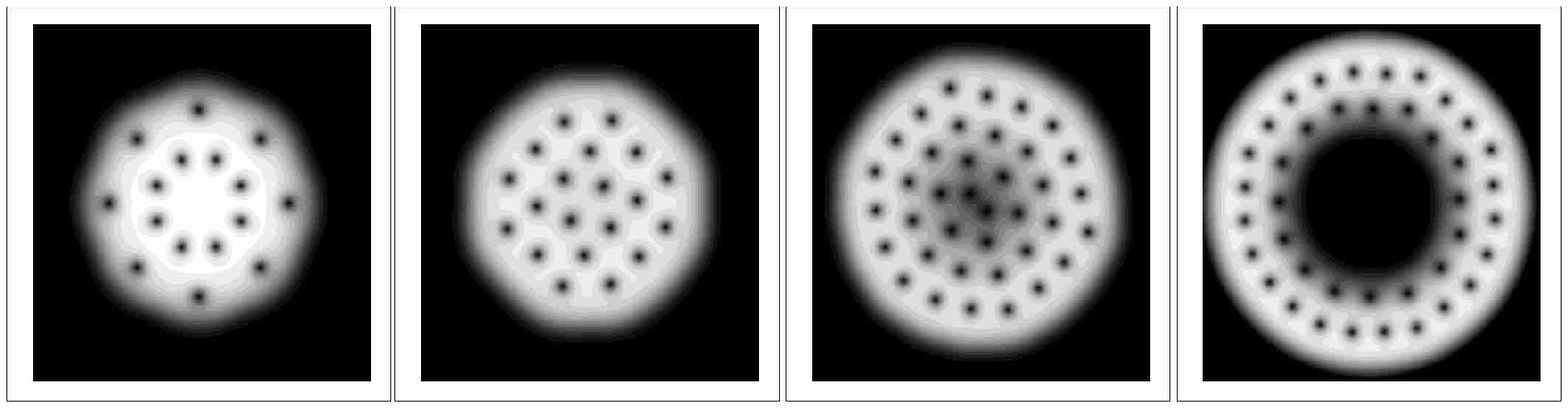}}
 \caption{($\alpha=0.9$) Top view of the condensate (up) and density contours in the plane  $z=0$ (down) for
  $\O/\o_\perp=0.32$ (a), $0.4$ (b), $0.48$ (c)
  and $0.56$ (d).} \label{fig-a09-top}
\end{figure}

\begin{figure}[!h]
\centerline{\includegraphics[width=0.9\columnwidth]{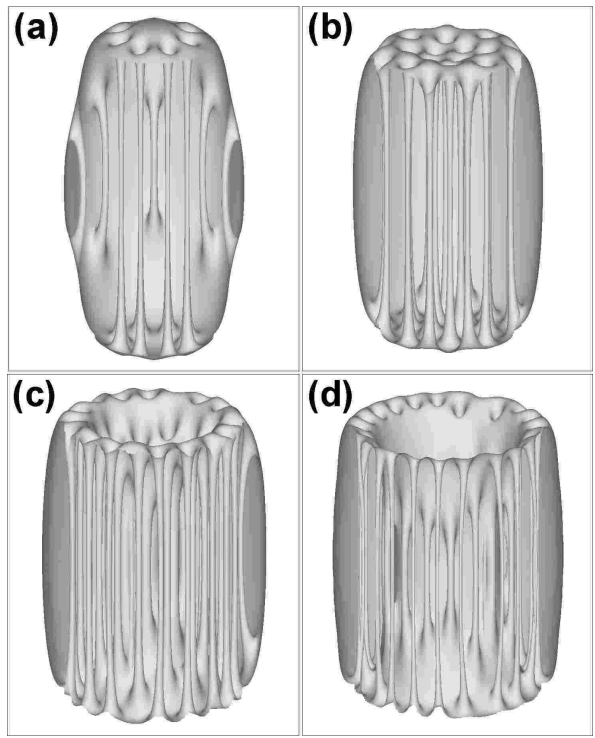}}
 \caption{($\alpha=0.9$) Side view of the condensate for $\O/\o_\perp=0.32$ (a), $0.4$ (b), $0.48$ (c)
  and $0.56$ (d).} \label{fig-a09-side}
\end{figure}

This is the case closest to the experiments \cite{BD}. The special feature of this case is that one has to achieve
 larger values of $\O$ in order to obtain giant vortices.
 The density profile of the solutions are shown in figures
  \ref{fig-a09-top} and
\ref{fig-a09-side}.

 Figure \ref{fig-a09-side} show the three dimensional structure of vortices.
There are isolated single quantized vortices, forming a
lattice. Increasing $\O$ leads to a denser lattice (20
vortices for $\O/\o_\perp=0.4$ and 38 for $\O/\o_\perp=0.48$). As
a consequence, the angular momentum (figure \ref{fig-lz}) grows
rapidly to high values.

>From $\O/\o_\perp=0.48$, the vortices near the center of the
condensate start to merge, leading to a central structure similar
to that displayed in figure \ref{fig3}(a). For $\O/\o_\perp\geq
0.56$, the central vortices have merged into a giant vortex. The
lattice still exist around. Similarly to the experiments, the hole
is obtained for large values of angular velocity ($\O/\o_\perp\ge
0.56$).

It is interesting to note from the side view of the condensate
(figure \ref{fig-a09-side}) that most vortices of the lattice are
straight, but some bent vortices (U shape) exist. The U vortices
are either connected to the outer boundary (bending outwards) of
the condensate (figure \ref{fig-a09-side}a,c) or to the giant
vortex (bending inwards) (figure \ref{fig-a09-side}d).

\section{Conclusion}

We have studied stable configurations of the Gross Pitaveskii
energy when the trapping potential is modified to include a
quartic minus a harmonic term.

For weak quartic potentials, the solution evolves from a vortex
lattice to a vortex array with hole when the angular velocity $\O$
is increased. For stronger quartic potentials, giant vortices are
obtained for lower $\O$,  at a stage where the lattice is not so
dense. The typical structure of vortices is to have a central
giant vortex with an outer circle of vortices around. We believe
that there should be a criterion depending on the radius of the
condensate and the radius of the annulus that should characterize
the final structure of the giant vortex: whether there is or not a
circle of vortices around the giant vortex and its precise
location.

The form of the potential considered in our simulations was
inspired from recent experiments \cite{BD}. We have checked that keeping the exponential
part of the potential  instead of its {\em quartic minus harmonic}
approximation does not change the qualitative behaviour of the
solutions. This suggests that if this situation could be achieved
experimentally, it would allow to observe giant vortices for lower
velocities than previously, that is before reaching the fast rotation regime.

\hfill

\noindent {\bf Acknowledgements:} We would like to acknowledge
stimulating discussions with V. Bretin.

\end{document}